\documentclass[preprint]{aastex631}

\usepackage{graphicx}
\shorttitle{Antarctic TianMu Survey Project}
\shortauthors{Niu et al.}
\graphicspath{{figures/}}

\begin{document}
\title{Antarctic TianMu Staring Observation Project II: Data reduction and preliminary results}

\author[0000-0001-5796-8010]{Hubiao Niu}
\affiliation{XinJiang Astronomical Observatory, Chinese Academy of Sciences,150 Science 1-Street, Urumqi, Xinjiang 830011, PR China}
\author[0000-0001-5245-0335]{Jing Zhong}
\affiliation{Shanghai Astronomical Observatory, Chinese Academy of Sciences, 80 Nandan Road, Shanghai 200030, PR China}
\author{Yu Zhang}
\affiliation{XinJiang Astronomical Observatory, Chinese Academy of Sciences,150 Science 1-Street, Urumqi, Xinjiang 830011, PR China}
\author{Jianchun Shi}
\affiliation{Shanghai Astronomical Observatory, Chinese Academy of Sciences, 80 Nandan Road, Shanghai 200030, PR China}
\author{Rui Rong}
\affiliation{School of Astronomy and Space Science, Nanjing University, 210093 Nanjing, PR China}
\author{Jinzhong Liu}
\affiliation{XinJiang Astronomical Observatory, Chinese Academy of Sciences,150 Science 1-Street, Urumqi, Xinjiang 830011, PR China}
\author{Shiyin Shen}
\affiliation{Shanghai Astronomical Observatory, Chinese Academy of Sciences, 80 Nandan Road, Shanghai 200030, PR China}
\author{Zhenghong Tang}
\affiliation{Shanghai Astronomical Observatory, Chinese Academy of Sciences, 80 Nandan Road, Shanghai 200030, PR China}
\affiliation{School of Astronomy and Space Science, University of Chinese Academy of Sciences, No. 19A, Yuquan Road, Beijing 100049, PR China}
\author{Dan Zhou}
\affiliation{Shanghai Astronomical Observatory, Chinese Academy of Sciences, 80 Nandan Road, Shanghai 200030, PR China}



\begin{abstract}
The Antarctic TianMu Staring Observation Program is a time-domain optical sky survey project carried out in Antarctica, capable of large sky coverage, high-cadence sampling, and long-period staring. It utilizes the exceptional observing conditions in Antarctica to conduct high-cadence time-domain sky surveys. At present, we have successfully developed an 18-cm aperture Antarctic TianMu prototype, which has been deployed at Zhongshan Station in Antarctica for two consecutive years of trouble-free observations, during which more than 300,000 original images were obtained. This paper systematically outlines the commissioning data of the prototype telescope in 2023, the primary data processing pipeline, and the preliminary data products. The core pipeline encompasses four key stages: Data preprocessing, instrumental effect correction, astrometric solution, and full-field stellar photometry. Here, we release the 2023 data products, which specifically include reduced image data and a photometric catalog, for which, preliminary analyses demonstrate robust performance. Using Gaia Data Release 3 as a reference catalog, the astrometric precision, quantified by the root mean square of positional errors, is determined to be better than approximately 2 arcseconds, validating the observational capabilities of the system. For a 30-second exposure, the detection limit in the G-band is achieved at 15.00~mag, with a detection threshold of 1.5~$\sigma$. The photometric errors are below 0.1~mag for the majority of stars brighter than 14.00~mag. Furthermore, it improves significantly, reaching better than 0.01~mag for most stars brighter than 11.00~mag and 12.00~mag when employing the adaptive aperture photometry and point spread function photometry methods, respectively.

\end{abstract}



\section{Introduction} 
\label{sect:intro}
As technology has advanced, astronomical observations have evolved from depicting a static universe to understanding a dynamic one. In key observation targets, such as supernovae, flare stars, active galactic nuclei, gravitational wave events, exoplanets, and compact objects, time-domain astronomy has made a series of major scientific discoveries and is emerging as a leading frontier in future international astronomical research\cite{2002SPIE.4836..154K,2009arXiv0912.0201L,2019PASP..131a8002B}. Based on multi-wavelength, continuous, and high-cadence observations, revealing the short-timescale variabilities of cosmic objects and exploring new types of celestial bodies and phenomena have become core goals in time-domain astronomy, to advance toward next-generation research\cite{2021AnABC..93..628L,2025RAA....25d4001H}.

At present, time-domain astronomical observations are developing in three directions: Shorter timescales, larger fields of view, and more sensitive detection capabilities. Existing ground-based time-domain astronomical observation equipment is located at mid-to-low latitudes, and single-aperture telescopes have a small Field of View (FoV), failing to balance the urgent needs for short timescales and continuous monitoring when observing transient sources\cite{2024FrASS..1104616H}.

The Antarctic TianMu Staring Observation Program (ATSOP) is intended to deploy multiple small-aperture ultra-large FoV telescopes in Antarctica, equipped with drift-scanning Charge-Coupled Device (CCD) cameras. Using image stacking for enhanced detection capability, it will conduct over 100 days of continuous high-cadence sampling across 1,200 square degrees of sky near the South Celestial Pole during each polar night, enabling observational research in short-timescale and long-period time-domain astronomy.

An 18-cm aperture Antarctic TianMu prototype (AT-Proto) has already been successfully developed and deployed. The prototype adopts a transmissive optical system design with an ultra-large FoV and a short focal length. Given that extremely low-temperature environments can easily cause malfunctions in telescope moving parts, cameras, optical imaging systems, and other components, we have employed an integrated temperature-controlled dome design. This design effectively regulates the temperature gradient inside the dome, ensuring that imaging variations caused by focal plane drift meet photometric requirements. Meanwhile, it significantly reduces the failure rate of electronic components, lowers resource demands for electronic heat dissipation and dome temperature control, and extends the telescope's operational lifespan. For the observation mode, a drift-scanning CCD camera is installed at the focal plane of the optical telescope, for use with staring observations. Long-duration integrated observation of faint celestial objects is achieved through charge tracking. This eliminates the need for a telescope servo motion mechanism, resulting in low development and operational costs as well as high reliability. This design is therefore more suitable for operation in the harsh Antarctic environment and is favorable for unattended operation.

By integrating a temperature-controlled dome with trackless drift-scanning technology, AT-Proto has withstood extreme Antarctic challenges, including subzero temperatures, frosting, long-term unattended operation, and limited energy resources. Deployed at Zhongshan Station in Antarctica for two consecutive years of trouble-free observation campaigns, it has recorded over 300,000 images.

\section{Observation}
\label{sect:obs}

AT-Proto was transported to Antarctica's Zhongshan Station using the Xue Long research vessel on October 31, 2022. Thanks to its integrated structural design, the entire system ensured exceptional convenience for both transportation and installation. Following ground foundation casting at Zhongshan Station, the equipment was hoisted into position, and the installation and debugging processes were swiftly completed through horizontal alignment and pointing adjustments. Commissioning observations commenced on February 7, 2023.

At Zhongshan Station, AT-Proto features a fixed horizon pointing of az~=~359.61$^{\deg}$ and alt~=~45.72$^{\deg}$. Given the station's geographic coordinates of 69$^{\deg}$22'24"S, 76$^{\deg}$22'40"E, the central declination of its observation field is Dec~=~-23.653$^{\deg}$. With a 9.5$^{\deg}$~×~9.5$^{\deg}$ FoV, AT-Proto can sample the observed sky region for approximately 40 minutes daily. It commenced formal observations on February 20, 2023, and transmitted observation images back to the ground. Analysis indicated that the Full Width at Half Maximum (FWHM) of point source images was 2 pixels, satisfying the intended requirements.

\begin{figure*}[!htb]  
   \centering
   \includegraphics[scale=0.7]{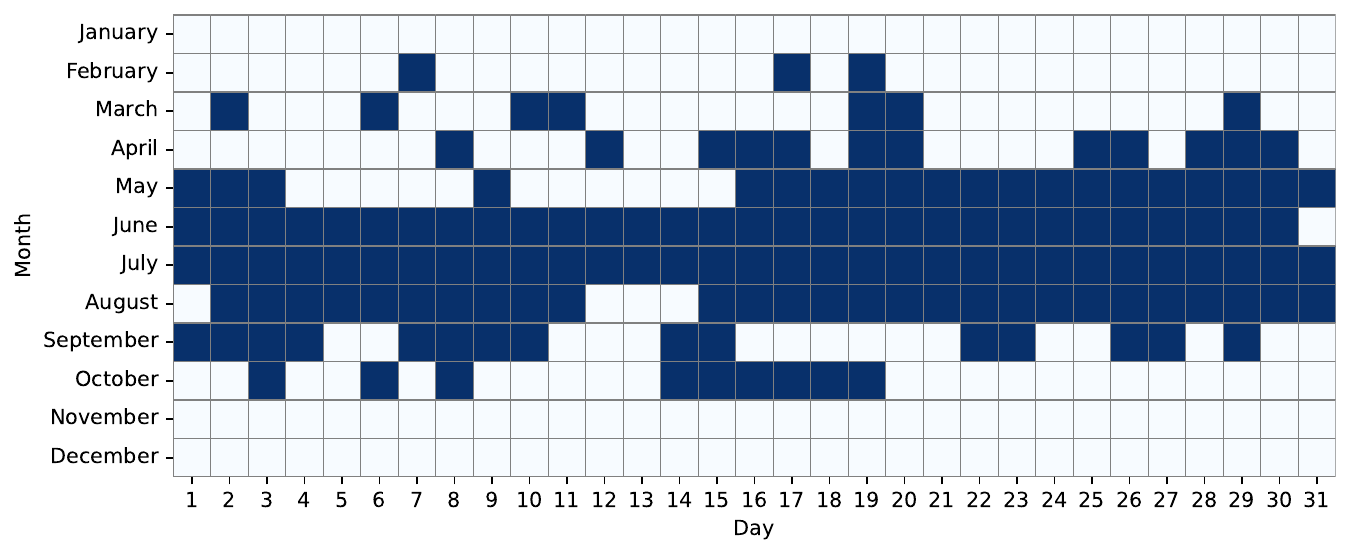}
    \caption{ Days with ATSOP observation data in 2023. The dark blue areas show the number of days observed. }
    \label{fig:1}
\end{figure*}

Since early March, drift-scan observations and tests have been conducted under favorable weather conditions, with a 30-second exposure time chosen for the sky survey. Additionally, a continuous observation test during polar night from May 21 to July 15 was successfully completed. Starting from the commissioning observation in February when the AT-Proto was powered on, the system has remained continuously operational without power interruption until October 26, achieving 248 consecutive days of fault-free operation. During this observation season, valid observation data were obtained over a total of 145 days (see in Figure~\ref{fig:1}), recording 161,805 observation images with a total data volume (including test data) of 3.35~TB.

We have conducted a preliminary assessment of the image quality and defined images using Analog-to-Digital Units (ADU) with a background sky brightness value less than 25,000~ADU and a number of detectable stars greater than 1,000 as high-quality images. In 145 total observation days, 79 days yielded more high-quality images than low-quality ones, representing 54\% of the total. During the 55-day polar night observation period, 46 days yielded high-quality images, accounting for 83\%. For the polar night period, most low-quality images were caused by excessively high background sky brightness from solar or lunar interference. Only 6 days of low-quality images were due to poor weather conditions, defined as conditions with a background median value below 25,000~ADU and fewer than 1,000 stars in the FoV.

\section{Data Reduction} 
\label{sec:reduction}

\begin{figure*}
    \begin{minipage}[t]{0.92\linewidth}  
    \centering
    \includegraphics[ width=1.05\linewidth, height=0.4\linewidth, keepaspectratio=False ]{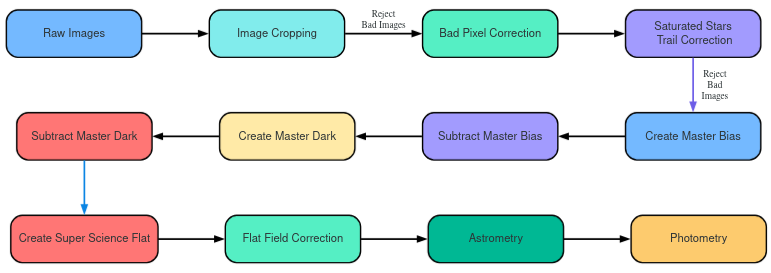} 
    \caption{Data reduction flowchart, showing the key processing steps from raw observations to final science-ready products. The pipeline incorporates parallel processing branches with quality control checks at each critical transition point.}
    \label{fig:flowchart}
    \end{minipage}
\end{figure*}

The AT-Proto data processing pipeline operates in parallel mode, with standardized steps performed sequentially: Image cropping, bad pixel correction, saturation-induced bleed trail correction, and systematic error calibration. The workflow diagram is shown in Figure~\ref{fig:flowchart}.

\subsection{Raw Data Organization} 
\label{ssec:raw} 

\begin{figure}
    \begin{minipage}[t]{0.99\linewidth}  
    \centering
    \includegraphics[ width=0.99\linewidth, height=0.7\linewidth, keepaspectratio=False ]{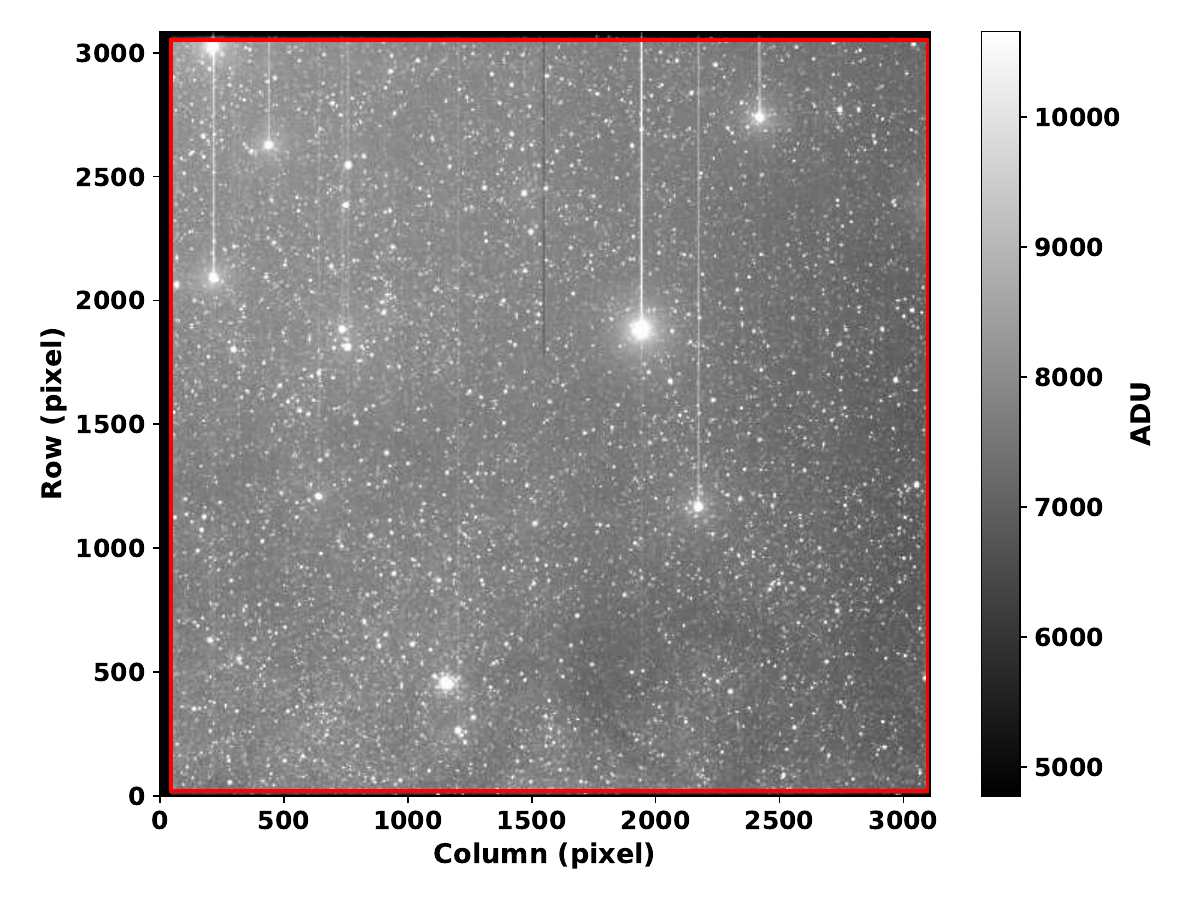} 
    \caption{ The original image and the region to be cropped, with the valid data area indicated by the red rectangular box.}
    \label{fig:cropregion}
    \end{minipage}
\end{figure}

The raw dataset consisted of 31 publicly available dark frames with exposure times spanning 1~--~180 seconds (6 exposures at 1 second each and five exposures at 5~s, 10~s, 50~s, 110~s, and 180~s), accompanied by daily observational data organized in “YYYY-MM-DD” directories containing bias frames and science images. All original 3108~×~3086 pixel images (without overscan regions) were uniformly cropped to eliminate edge flux anomalies using slice operations [21:3050, 45:3101], with the valid data region demarcated by the red rectangle in Figure~\ref{fig:cropregion}. The processed Flexible Image Transport System (FITS) files retained standard headers including DATE-OBS, EXPOSURE ($\sim$30~s), and instrument parameters for pipeline processing, while the cropped dimensions represented the scientifically validated FoV. Following image cropping, we performed preliminary quality control by flagging and rejecting frames with background values exceeding 20,000~ADU as potentially compromised quality images. This conservative threshold was empirically determined to efficiently identify and exclude the majority of problematic exposures while preserving scientifically useful data for subsequent analysis. Finally, the celestial coordinates of the image center were computed by transforming the telescope’s pointing direction from the local alt-azimuth frame to the International Celestial Reference System (ICRS) Using the observation timestamp (combined from DATE-OBS and TIME-OBS header information) and the observatory’s geographic coordinates, we applied the coordinate transformation pipeline from the astropy Python package to convert the telescope’s (az, alt) pointing to equatorial coordinates (RA, Dec)\citep{astropy:2022}. These values were then written to the FITS header as the reference point (CRVAL) for the World Coordinate System (WCS), along with the tangent-plane projection parameters (CTYPE) and pixel scale information (CDELT). The WCS solution is centered on the image midpoint (CRPIX) to establish the astrometric solution for subsequent analysis. The results are recorded in the FITS header in standard format.

\subsection{Detector Artifact Correction}
\label{ssec:artifacts}

\begin{figure}
    \begin{minipage}[t]{0.99\linewidth}  
    \centering
    \includegraphics[width=0.99\textwidth, keepaspectratio=True, angle=0,scale=1]{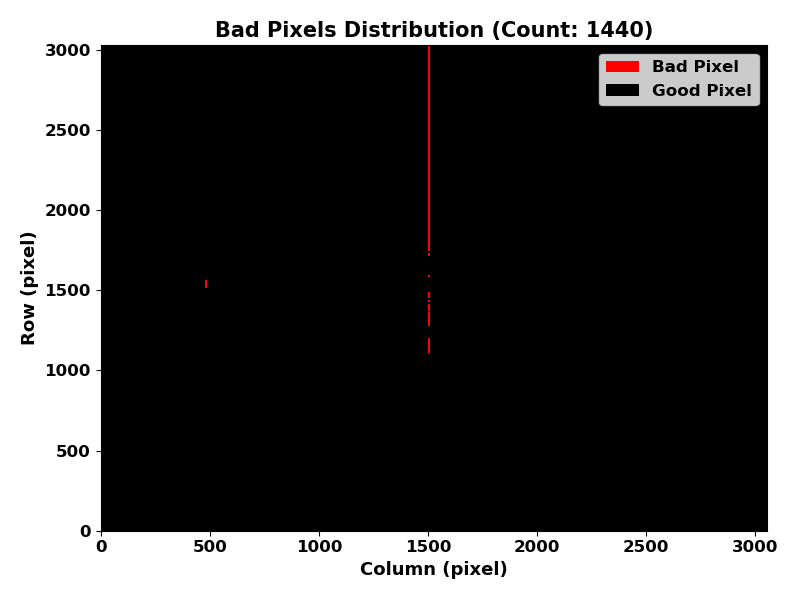} 
    \caption{Mask of bad pixels, derived from different epochs of super master sky flats.} 
    \label{fig:ccdmask}
    \end{minipage}
\end{figure}

The master bad pixel mask was constructed with a multi-epoch validation approach using super sky flats taken from April to October 2023. We randomly selected four representative monthly datasets (June, July, September, and October) to identify column defects and bad pixels. The final mask incorporated only pixels consistently flagged across all epochs, ensuring robustness against transient artifacts. Defects were identified using \textit{ccdproc.ccdmask} with $7 \sigma$ clipping on the median of super science flat frames. Figure~\ref{fig:ccdmask} shows the spatial distribution of defective pixels across the CCD, with red markers identifying bad pixels in the color-coded defect map. All affected pixels were corrected with bi-variate spline interpolation, using the \texttt{RectBivariateSpline} function, which performed cubic spline interpolation while preserving local gradient continuity. This approach showed superior performance compared with nearest-neighbor methods, particularly for clusters of defective pixels. For saturation trails along the readout direction, we implemented a multi-stage correction, identification of saturated stars using morphological criteria:
\begin{itemize}
\item Minimum pixel count: $\geq$4
\item Maximum pixel count: $\leq$324
\item Eccentricity: $\leq$0.7
\end{itemize}

\begin{figure}[htbp]
\centering
\begin{minipage}[t]{0.99\linewidth} 
\includegraphics[height=0.5\textwidth, width=0.48\textwidth, trim=0 0 0cm 0, clip]{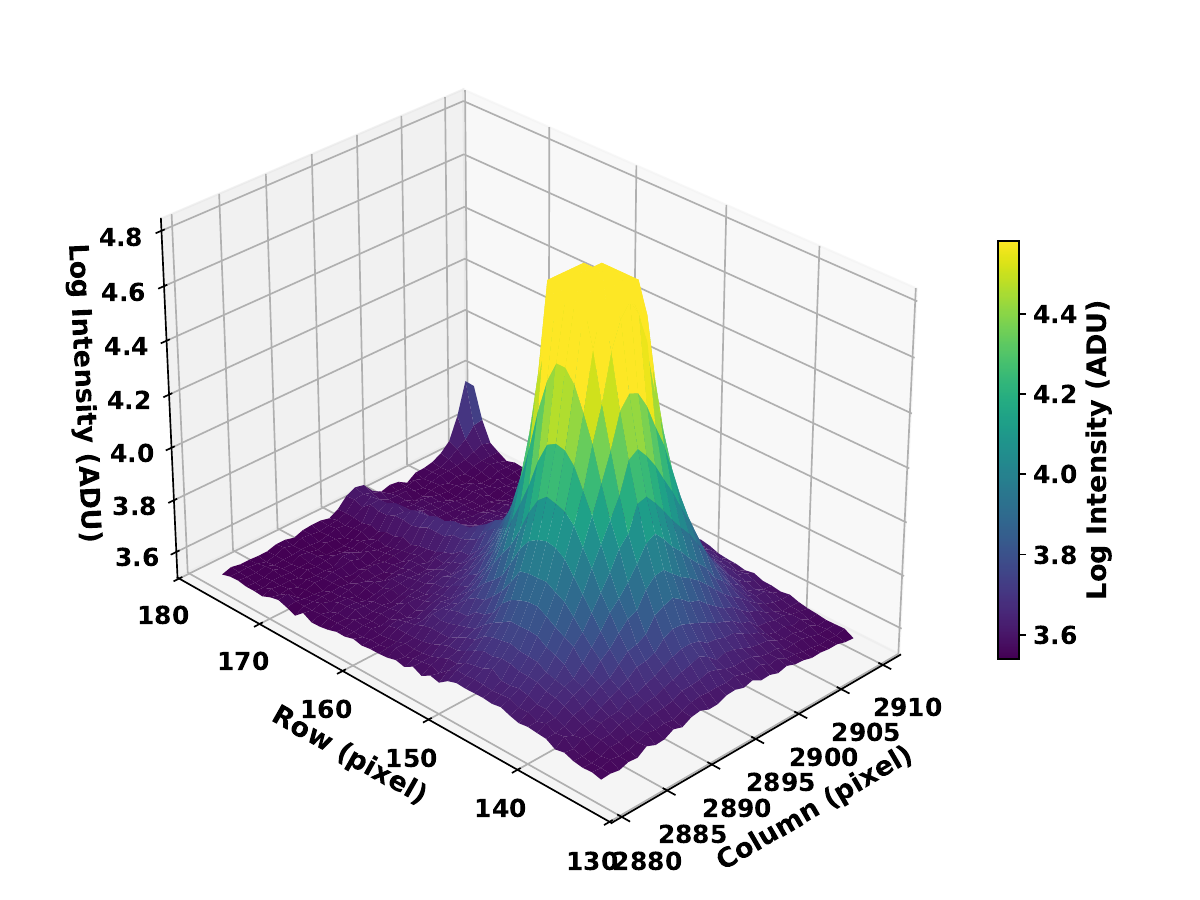} 
\includegraphics[width=0.48\textwidth, trim=0.8cm 0 0 0, clip]{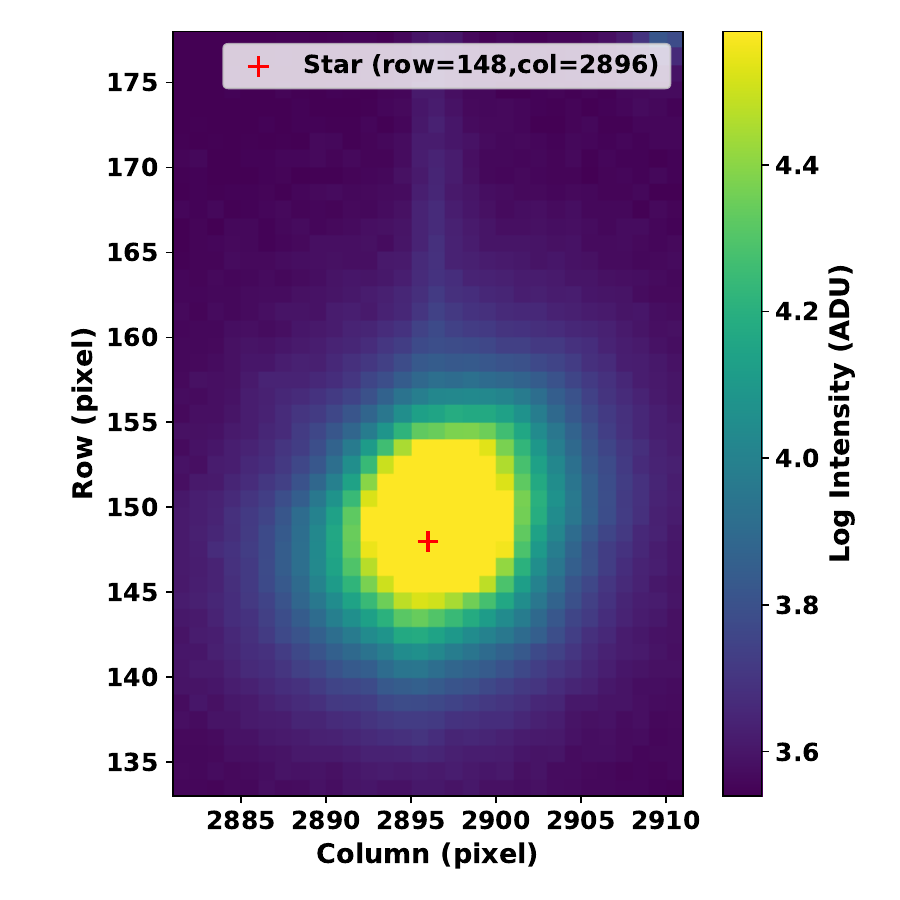} 
\caption{Characterization of a saturated star’s profile with charge bleeding artifacts. Left panel: Three-dimensional surface plot showing the star’s PSF with a prominent vertical trench (columns 2893-2899) caused by charge blooming. Right panel: Two-dimensional intensity profile, showing the dark bleeding trail extending from the saturated core.}
\label{Fig:trail_profile}
\end{minipage}
\end{figure}

\begin{figure}[htbp]
\centering
\begin{minipage}[t]{0.99\linewidth} 
\includegraphics[width=0.5\textwidth]{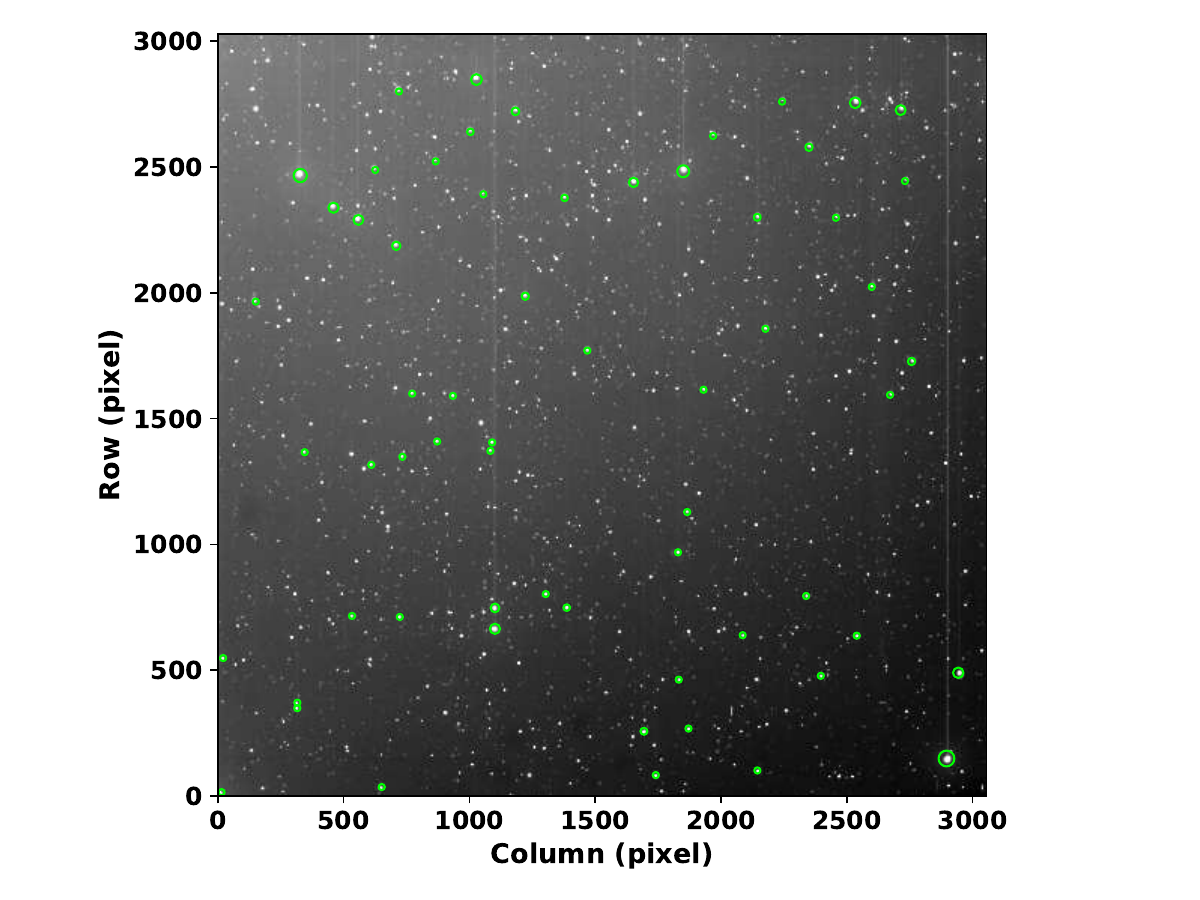} 
\includegraphics[width=0.5\textwidth]{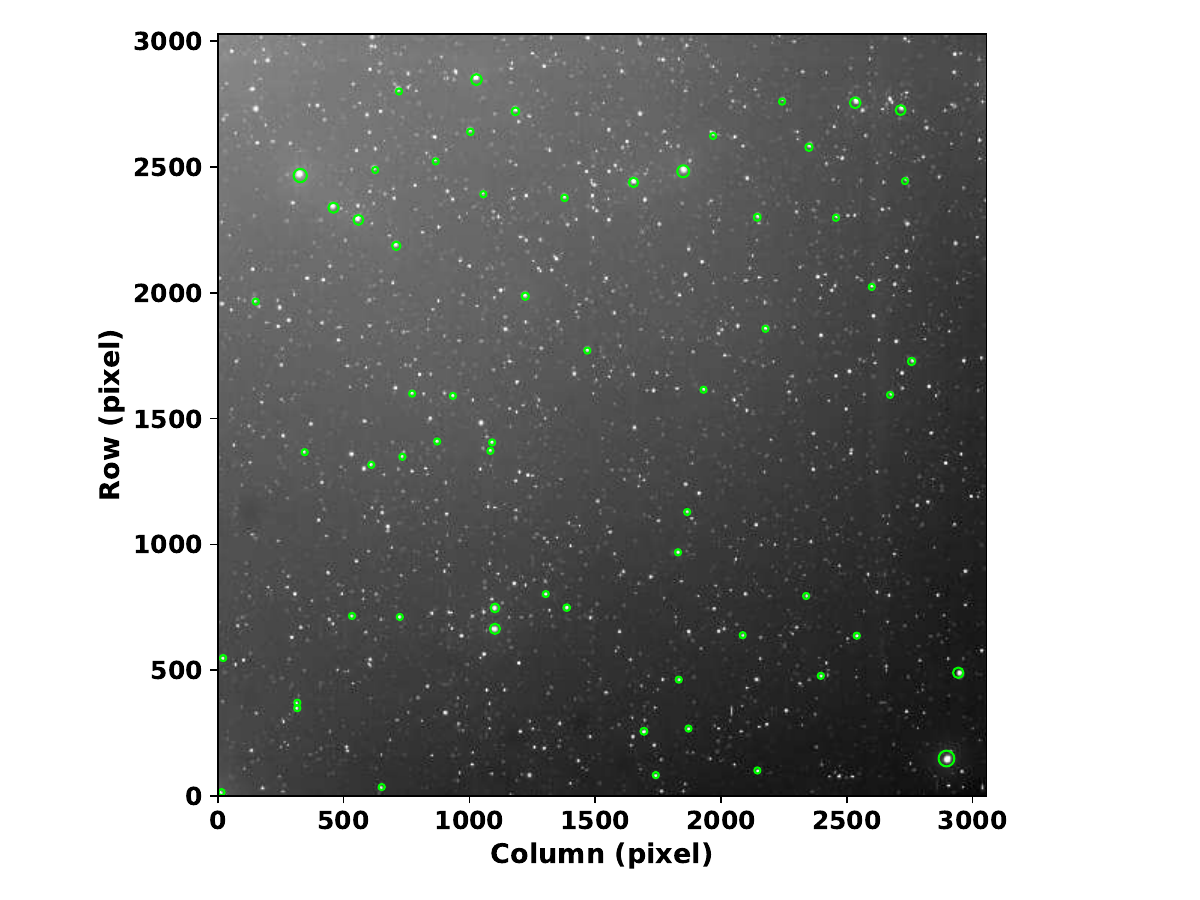} 
\caption{Comparison of raw and processed images, demonstrating charge bleeding correction. Left panel: Original image showing vertical bleeding artifacts from saturated stars. Right panel: Result after applying our correction algorithm, effectively removing the vertical streaks while preserving the original stellar profiles. }
\label{Fig:trail_corr}
\end{minipage}
\end{figure}

\begin{figure*}[htbp]
\centering
\begin{minipage}[t]{0.99\linewidth} 
\includegraphics[width=0.95\textwidth]{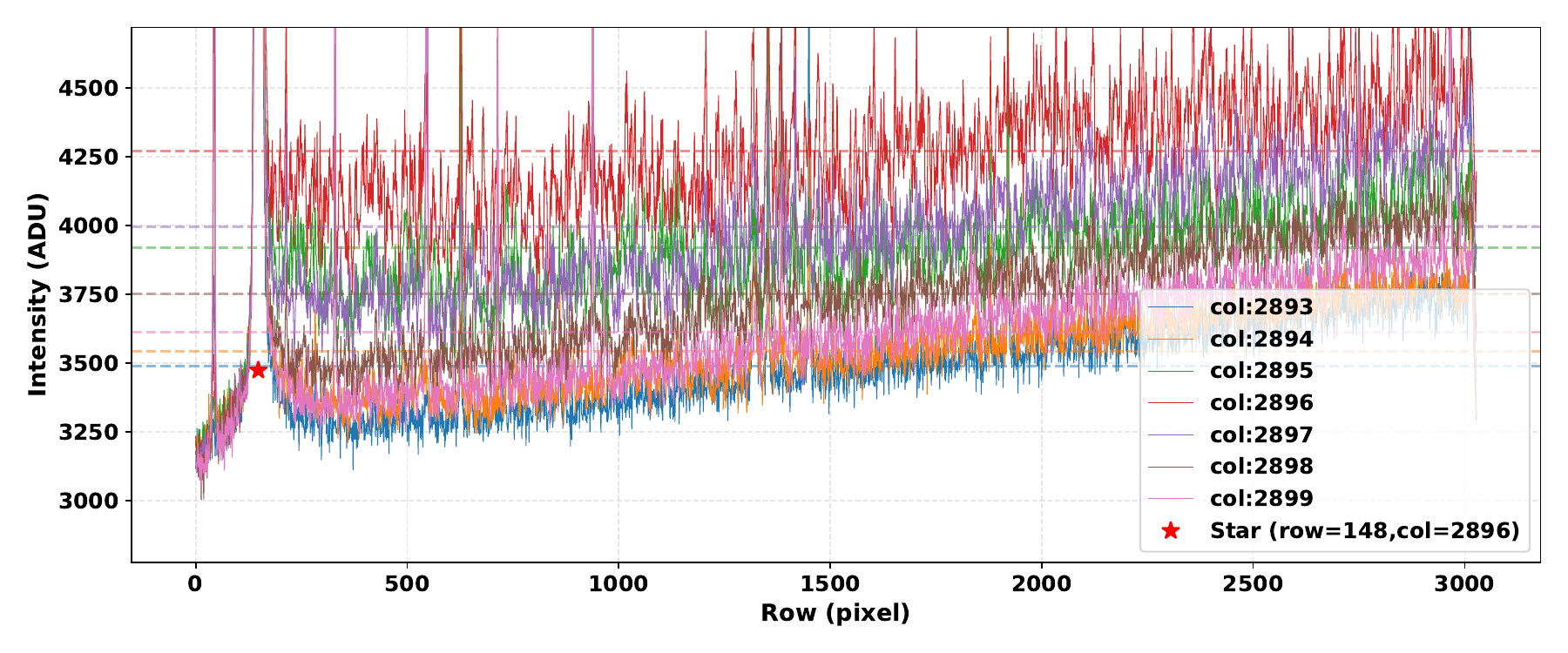} 
\includegraphics[width=0.95\textwidth]{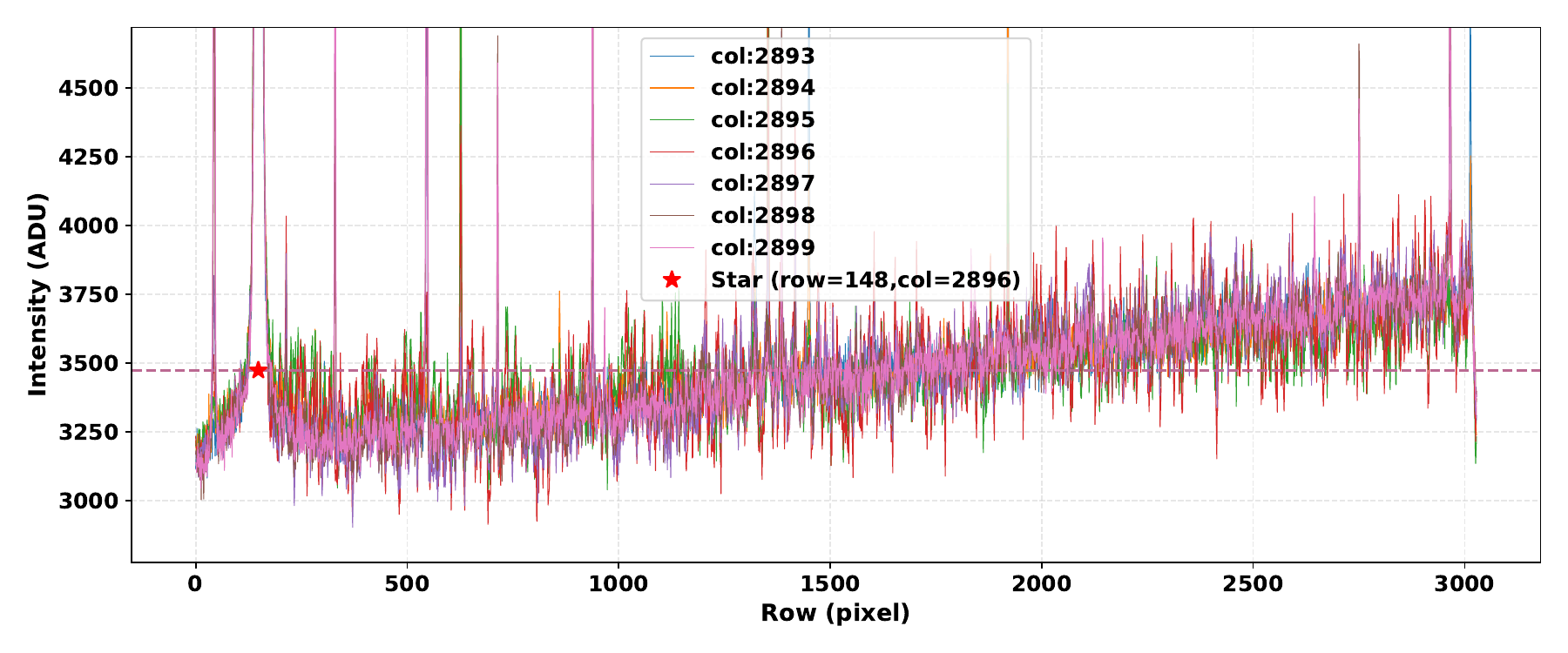} 
\caption{Charge bleeding correction across columns 1766-1772. Top panel: Uncorrected column profiles showing significant saturation bleeding extending vertically from the stellar core (column 1769). Bottom panel: Calibrated profiles after artifact removal, with affected columns returning to the median background level. The dashed lines represent the median value of corresponding columns. }
\label{Fig:trail_columns}
\end{minipage}
\end{figure*}

Since our observations were conducted without a mechanical shutter, bright stars saturate and generate charge overflow trails along the readout direction (i.e., blooming, see Figure~\ref{Fig:trail_profile}). To correct for these trails, we first identify saturated sources ($ADU > 65000$) using morphological criteria: Pixel count range $4 \leq N_{\mathrm{pix}} \leq 500$, eccentricity $e \leq 0.8$, and verification through $r = 5$ pixel aperture photometry. For each confirmed saturated star at position $(x,y)$, we define the trail region extending vertically from $y+r$ to $y_{\mathrm{max}}$ (the detector edge) with a horizontal width $w = \min(2D_{\mathrm{eq}}, 8)$, where $D_{\mathrm{eq}}$ is the equivalent diameter of the saturated core. The background level $F_{\mathrm{bkg}}$ is estimated using a robust sigma-clipping algorithm combined with the \texttt{SExtractorBackground} estimator. Specifically, a two-dimensional background model is constructed from the raw image using the \texttt{Background2D} class, with a box size of $50 \times 50$ pixels and a filter size of $3 \times 3$ pixels. This approach ensures that the background estimation is insensitive to outliers caused by bright sources and detector artifacts. For each column, $j$, within the trail region $[x-w/2, x+w/2]$, the trail profile is independently fitted using a Locally Weighted Scatterplot Smoothing (\texttt{LOWESS}) regression. The \texttt{LOWESS} algorithm, with a smoothing parameter of $0.5$, is applied to the valid pixels in each column, where valid pixels are defined as those within $3\sigma$ of the median value after sigma clipping. The fitted trail profile is then interpolated to the whole column $y' \in [y+r, y_{\mathrm{max}}]$ using a quadratic interpolation function. The corrected flux, $F^{\mathrm{corr}}$, for each pixel $(j, y')$ is calculated as
\begin{equation}
F^{\mathrm{corr}}(j, y') = F^{\mathrm{raw}}(j, y') - F^{\mathrm{trail}}(j, y') + F_{\mathrm{bkg}},
\end{equation}
where $F^{\mathrm{trail}}(j, y')$ is the fitted trail profile and $\langle F_{\mathrm{bkg}} \rangle$ is the local background value from the two-dimensional background model. This method effectively removes bleed artifacts caused by saturated sources while accurately restoring the flux of stars contaminated by the trails (see Figure~\ref{Fig:trail_corr} and Figure~\ref{Fig:trail_columns}).

\subsection{Basic Calibration} 
\label{ssec:basic}

The data reduction pipeline follows standard astronomical calibration procedures, with detector-specific optimizations \cite{2017zndo...1069648C}. All calibration frames are acquired during the same observing run under identical detector configurations, except for the dark frames, which are taken only once.

\paragraph{Bias correction} utilizes all available zero-second exposure frames ($N_{\mathrm{bias}}$) acquired during the observing run. The master bias is constructed through $\pm3\sigma$-clipped median combination to minimize readout noise. Each science frame $F_{\mathrm{raw}}$ then subtracts the master bias frame:

\begin{equation}
F_{\mathrm{bias}}= F_{\mathrm{raw}} - B_{\mathrm{master}}.
\end{equation}

\paragraph{Dark calibration} begins with bias correction of individual dark frames $D_{\mathrm{raw}}^{(i)}(t_{\mathrm{dark}})$ via $D_{\mathrm{bias}}^{(i)} = D_{\mathrm{raw}}^{(i)} - B_{\mathrm{master}}$, followed by creation of exposure-time-specific master darks through $\pm3\sigma$-clipped median combination of $N_{\mathrm{dark}}$ frames (typically 5-6 per time bin). Science frames $F_{\mathrm{bias}}(t_{\mathrm{exp}})$ are matched to the nearest valid master dark $D_{\mathrm{master}}(t_{\mathrm{dark}})$, with scaled subtraction:  
\begin{equation}
F_{\mathrm{dark}} = F_{\mathrm{bias}} - \left(\frac{t_{\mathrm{exp}}}{t_{\mathrm{dark}}}\right)D_{\mathrm{master}}(t_{\mathrm{dark}}).
\end{equation}

\paragraph{Flat fielding} uses a super science flat constructed exclusively from science exposures. The master flat is generated through a multi-step process, applied to 100 randomly selected science frames that have been previously bias- and dark-corrected. The construction begins with robust background estimation for each frame and its standard deviation, with $\sigma$ clipping to exclude outliers. All pixels deviating by more than $5\sigma$ from their median value (primarily stars) are masked as $np.NaN$ and excluded from further processing. The masked frames are then combined using a $3\sigma$-clipped median algorithm. This stacking method effectively suppresses residual contamination while preserving the true flat-field structure. The resulting super science flat is normalized to preserve the relative flux scaling of the original data. The final calibration of science frames is performed by dividing each science frame by the normalized master flat.  Key advantages of this process include compensation for the lack of dedicated flat-field frames, effective removal of stellar contamination, and preservation of the original flux scaling.

\begin{figure*}[htbp]
\centering
\begin{minipage}[t]{0.99\linewidth} %
\includegraphics[width=0.48\textwidth]{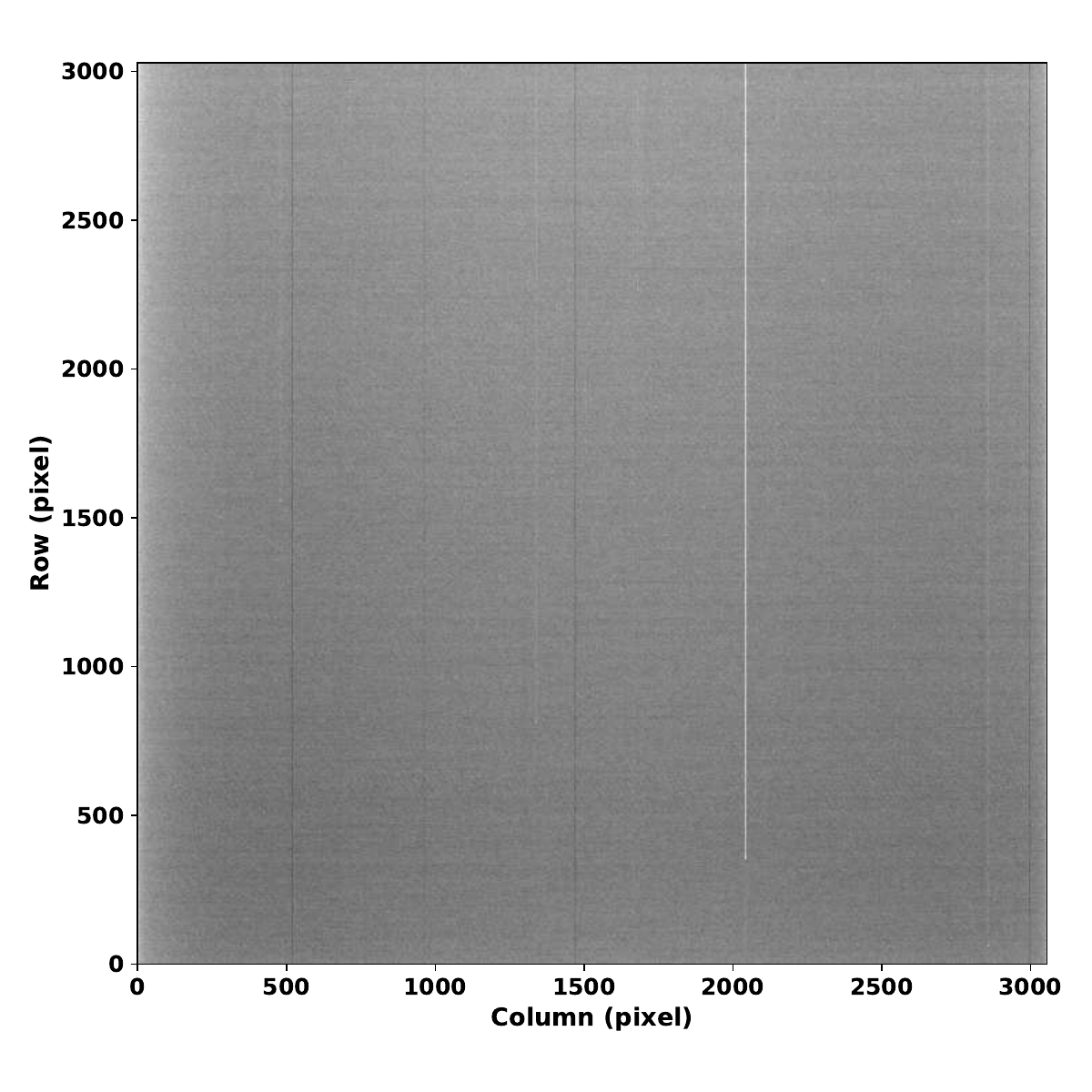} 
\includegraphics[width=0.48\textwidth]{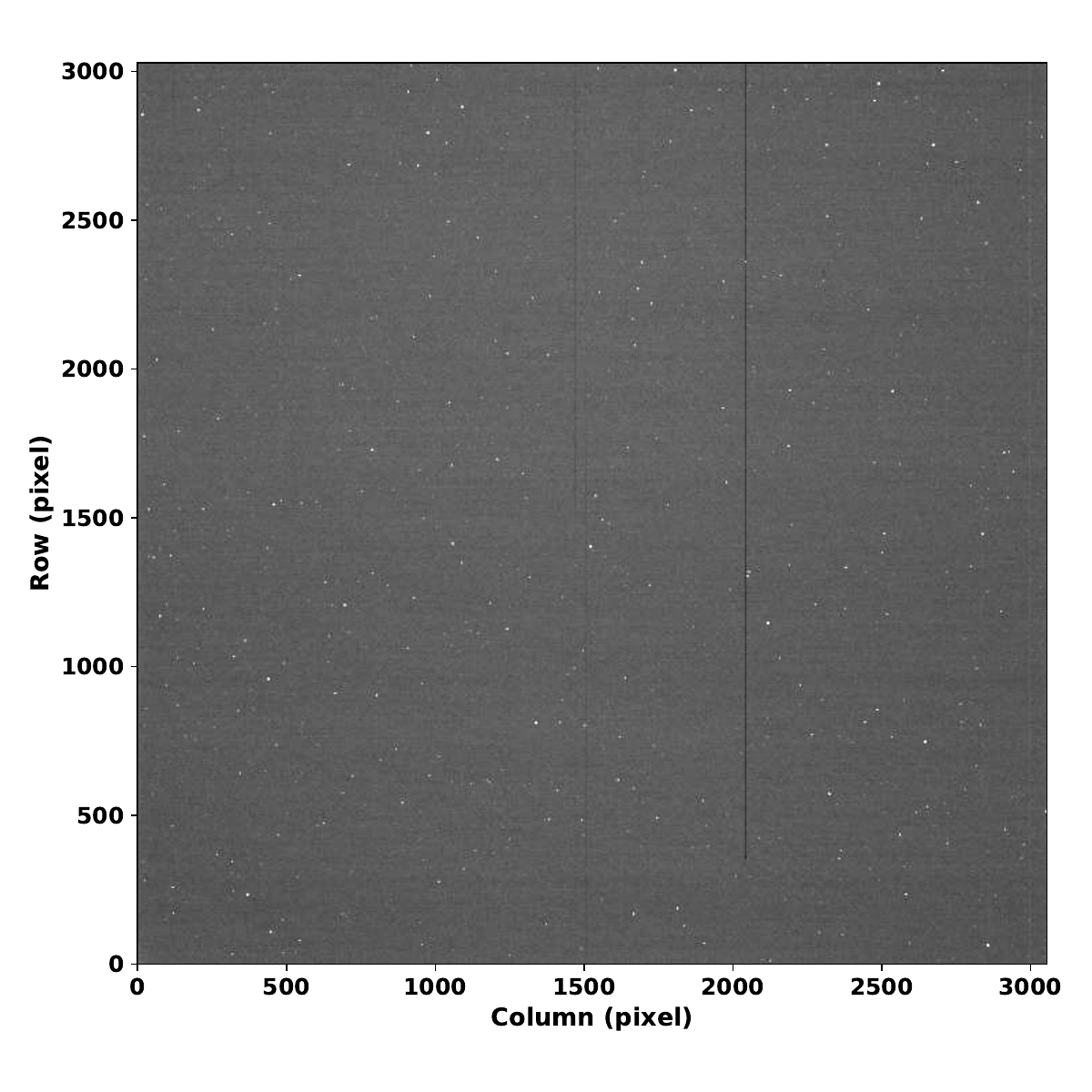} \\
\includegraphics[width=0.48\textwidth]{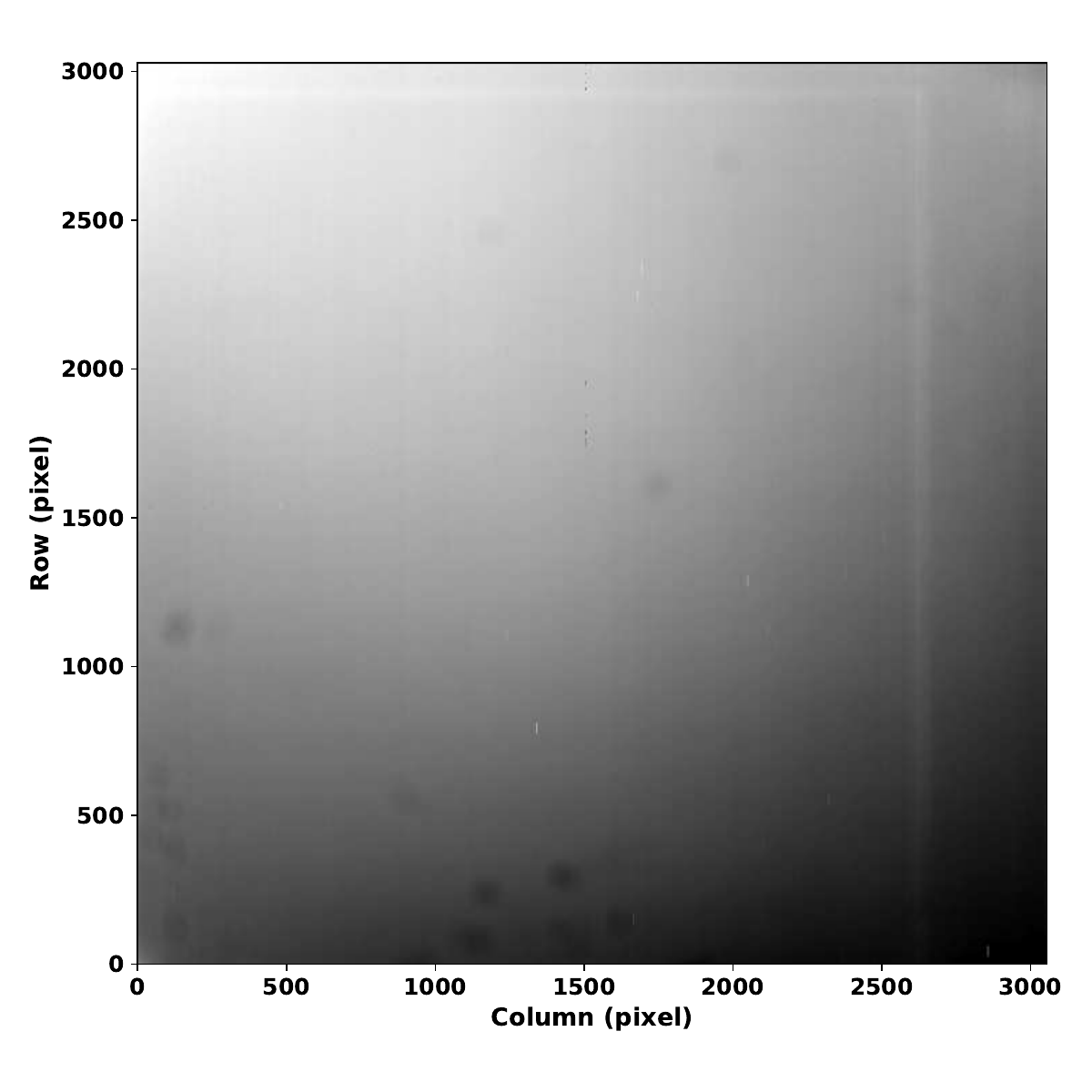} 
\includegraphics[width=0.48\textwidth]{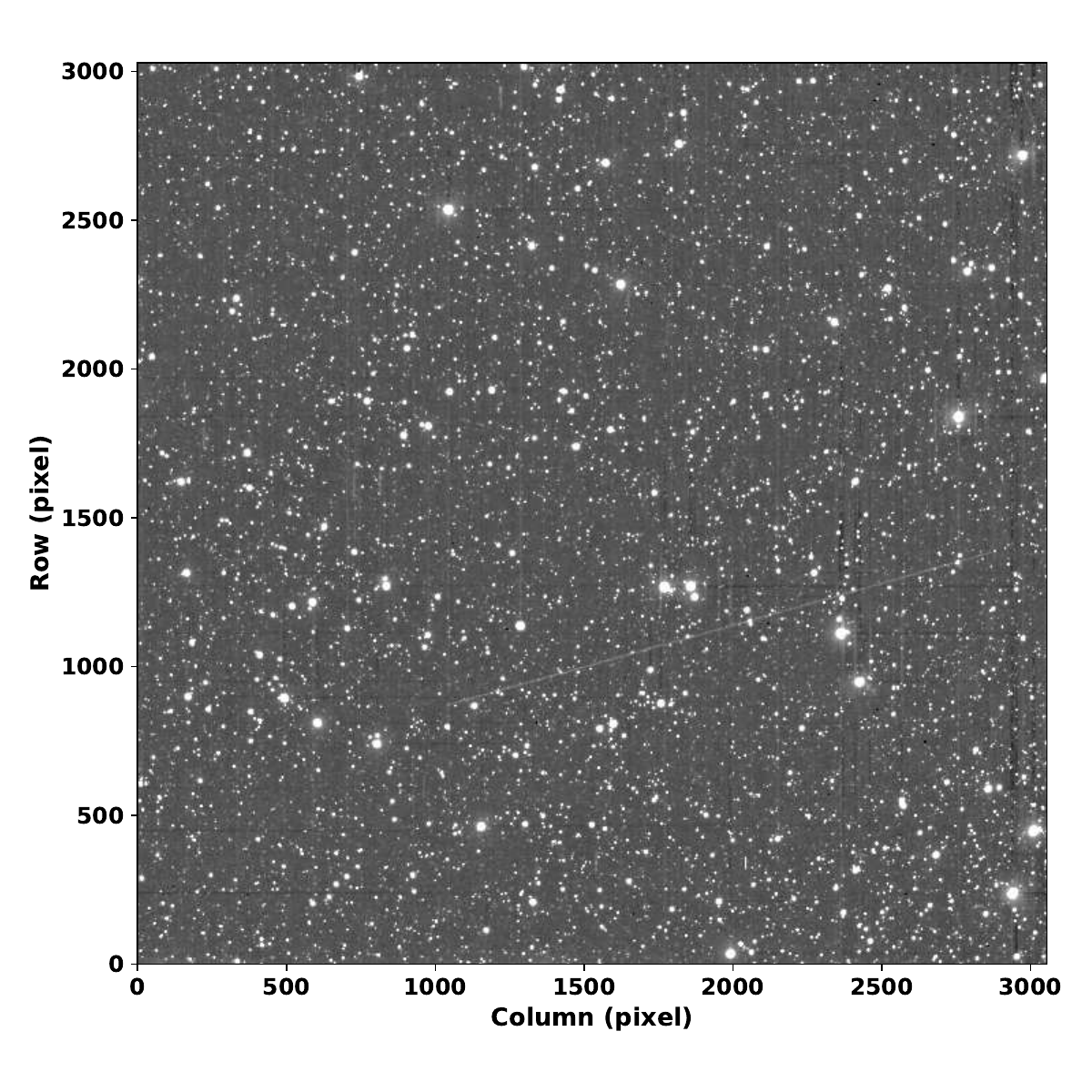} 
\caption{Calibration frame sequence and the final calibrated science frame. Top left panel: Master bias frame. Top right panel: Master dark frame for 50-second exposures. Bottom left panel: Super science flat field constructed from processed science frames. Bottom right panel: Final science frame after full calibration and background subtraction.}
\label{Fig:cal_frames}
\end{minipage}
\end{figure*}

\begin{figure}[htbp]
\centering
\begin{minipage}[t]{0.99\linewidth} %
\includegraphics[width=0.95\textwidth]{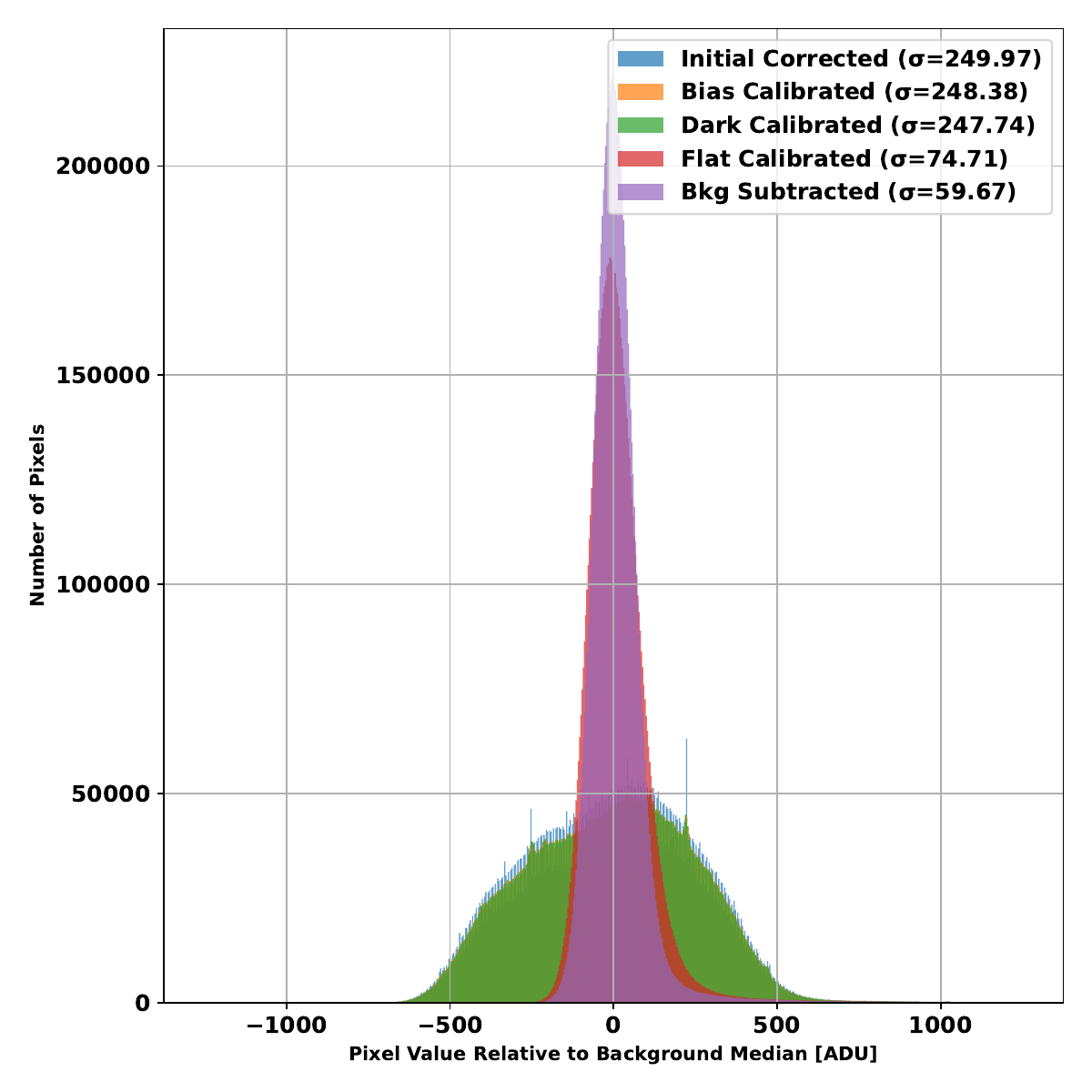} 
\caption{Evolution of pixel value distributions through the calibration process. The histogram sequence shows progressive refinement from raw science frames (blue) to bias-subtracted (yellow), dark-corrected (green), flat-fielded (red), and finally background-subtracted (purple) data.}
\label{Fig:cal_dist}
\end{minipage}
\end{figure}

 \paragraph{Precise sky background} estimation and removal is the final processing step using an optimized \texttt{Background2D} algorithm with SExtractor’s background estimator. The fitted two-dimensional background model is subtracted to eliminate sky brightness gradients while preserving source fluxes. The entire pipeline executes as a fully automated and parallelized workflow. Key calibration frames (including the master bias, master dark, super science flat, and final calibrated science frame) are shown in Figure~\ref{Fig:cal_frames}, while Figure~\ref{Fig:cal_dist} shows quantified progressive improvements in pixel-value distributions after each processing stage. Each processing step reduces the distribution width, with the standard deviation ($\sigma$) systematically decreasing as the calibration progresses.

\subsection{Astrometry Solution} \label{ssec:wcs}

\begin{figure*}[htbp]
    \centering
    \includegraphics[height=0.45\linewidth, width=0.45\linewidth]{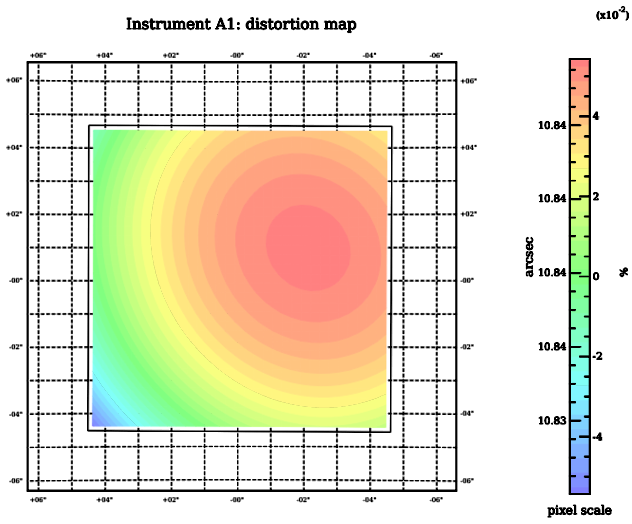}
     \includegraphics[height=0.45\linewidth, width=0.45\linewidth]{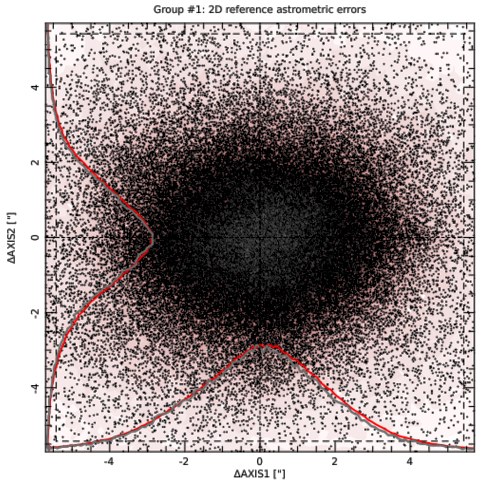}
    \caption{Left panel: Distortion map showing the spatial variation of pixel scale (arcsec/pixel) across the FoV after 3rd-order polynomial calibration with SCAMP. The color scale (red to blue) shows a nearly radially symmetric pattern with a symmetric center slightly offset towards the upper-right, where pixel scales decrease from 10.84~arcsec/pixel at the central region to 10.83~arcsec/pixel at the outer region. Right panel: Two-dimensional distribution of astrometric errors between the coordinates of detections and Gaia DR3 reference stars. The light red density map represents all sources, while black points mark stars with high S/N (\textgreater 100). Marginal distributions along both axes show the corresponding one-dimensional error distributions (red: all sources; gray: high S/N subset).
}
    \label{fig:distort_errors2d}
\end{figure*}

\begin{figure*}[htbp]
    \centering
     \includegraphics[height=0.45\linewidth, width=0.45\linewidth]{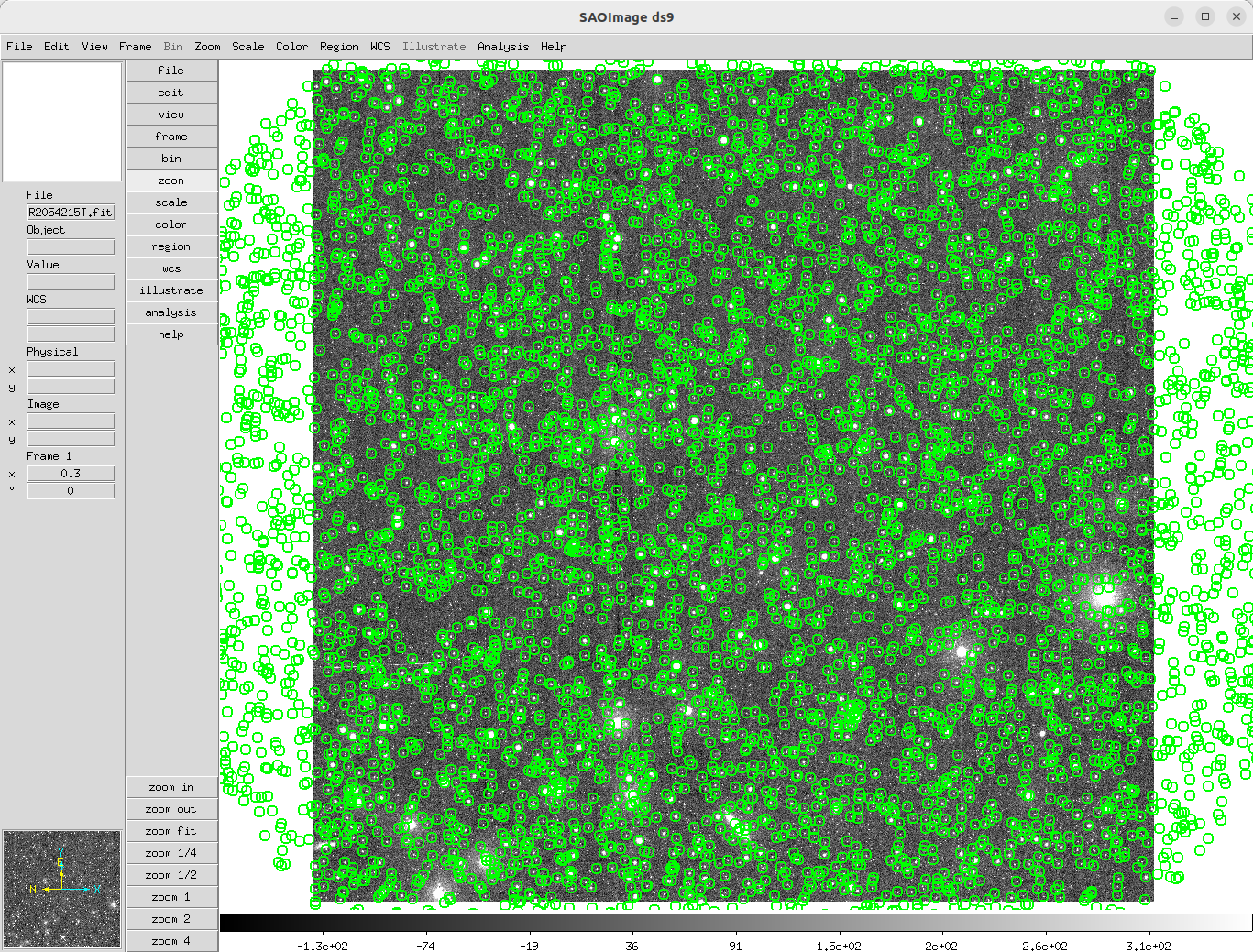}
     \includegraphics[height=0.45\linewidth, width=0.45\linewidth]{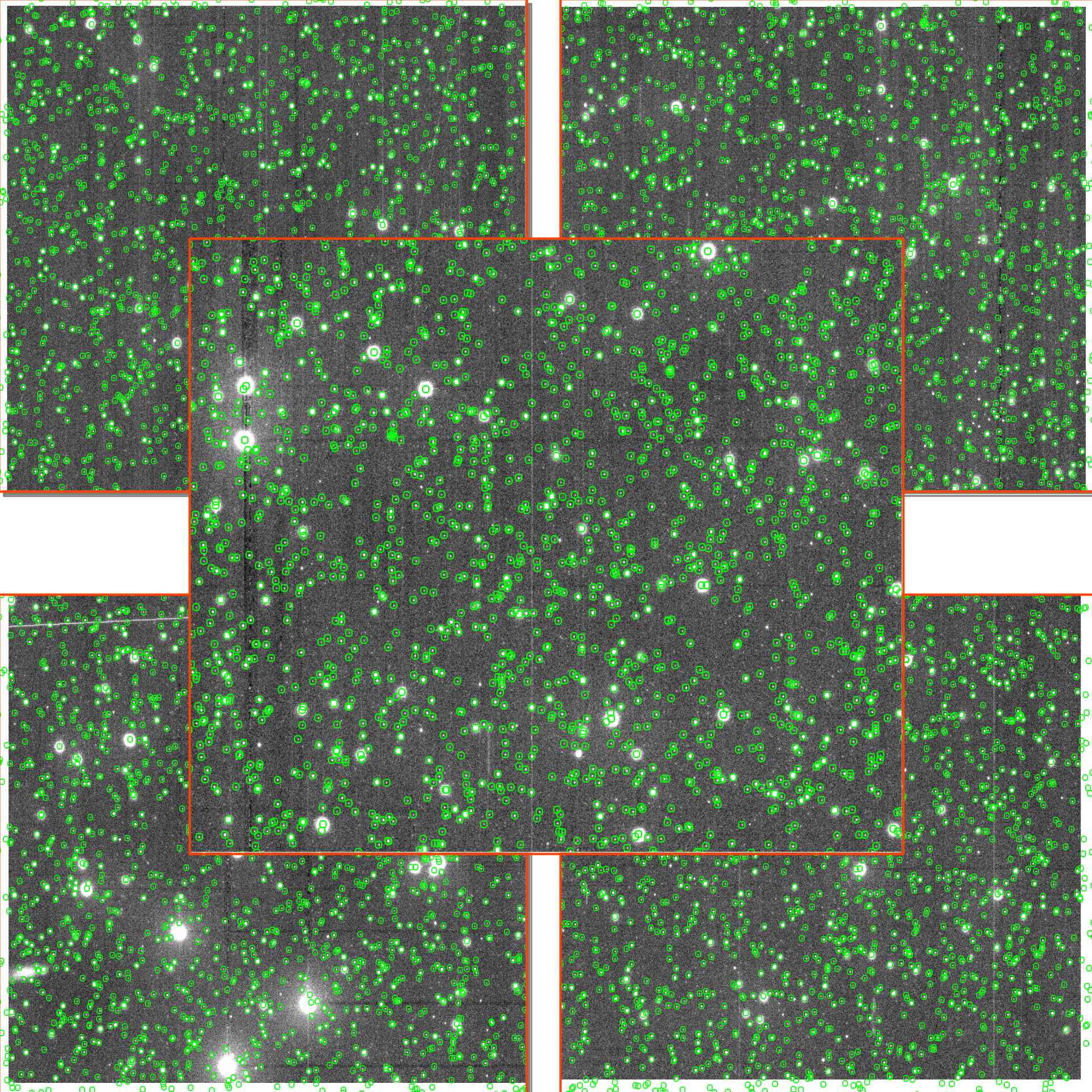}
    \caption{Astrometric alignment verification between the observed field and Gaia DR3 catalog. Left panel: Full-field match with Gaia DR3 reference positions, shown as empty green circles. Right panel: Magnified view of central and corner regions, demonstrating excellent alignment.}
    \label{fig:wcs}
\end{figure*}

Accurate astrometric calibration is crucial for correcting the geometric distortions inherent to wide-field imaging systems and obtaining high-precision celestial positions. Based on the preliminary WCS solutions derived from image center coordinates (RA, DEC) in section~\ref{ssec:raw}, the astrometric calibration was refined using the SCAMP\citep{2006ASPC..351..112B} with third-order polynomials, to address optical aberrations (including field curvature and coma) and detector non-planarity effects. Gaia Data Release 3 (DR3) served as the reference catalog for pattern matching online. The matching maximum scale-factor uncertainty was set to 20\%  with a $\geq$ requirement of 6 reference-stars for each frame. Given the aperture characteristics and spatial resolution of this telescope, we adopted a matching radius of 11~arcsec for cross-identification, and selected sources within the Gaia G-band magnitude brighter than 16~mag. The left panel of Figure~\ref{fig:distort_errors2d} shows the distortion map derived from SCAMP’s third-order polynomial solution, where the color-coded pixel scale shows a characteristic radial pattern. The systematic radial variation from 10.84~arcsec/pixel, at the field center, to 10.83~arcsec/pixel at the outer region, with a symmetric center slightly offset towards the upper-right, indicates the presence of a symmetric optical distortion pattern, which has been successfully modeled by polynomial fitting. The astrometric precision, as quantified by the RMS of position errors relative to Gaia DR3 reference stars, is measured to be approximately 1.9 and 1.7 arcsec in RA and Dec, respectively, demonstrating consistency with the reference catalog (see the right panel of Figure~\ref{fig:distort_errors2d}). The two-dimensional error distribution shows that the $1\sigma$ region achieves a precision better than 2 arcsec, confirming that our astrometric solution successfully accounts for field-dependent systematic errors. Figure~\ref{fig:wcs} shows the astrometric alignment between WCS-calibrated sources and Gaia DR3 reference positions, showing a close match even at the field edges.

\subsection{Photometry} \label{ssec:phot}

We use tools from PSF Extractor (PSFEx) to extract profiles of Point Spread Functions (PSFs)\citep{1996A&AS..117..393B}, then use the SExtractor software to perform photometry, including adaptive aperture photometry (\texttt{MAG\_AUTO}) and PSF photometry (\texttt{MAG\_PSF})\citep{2011ASPC..442..435B}. This photometry is performed with a detection threshold of $1.5 \sigma$, resulting in approximately 10,000 to 20,000 stars detected per image. The adaptive aperture photometry (\texttt{MAG\_AUTO}) and PSF photometry (\texttt{MAG\_PSF}) are used to measure the magnitudes of the detected stars. The resulting photometric catalog includes detailed information such as positions, fluxes, magnitudes, errors, and morphological parameters for each star. This catalog is saved in FITS\_LDAC format, which is compatible with various astronomical software tools for further analysis.

\begin{figure*}[htbp]  
    \centering
    \includegraphics[height=0.45\linewidth, width=0.45\linewidth]{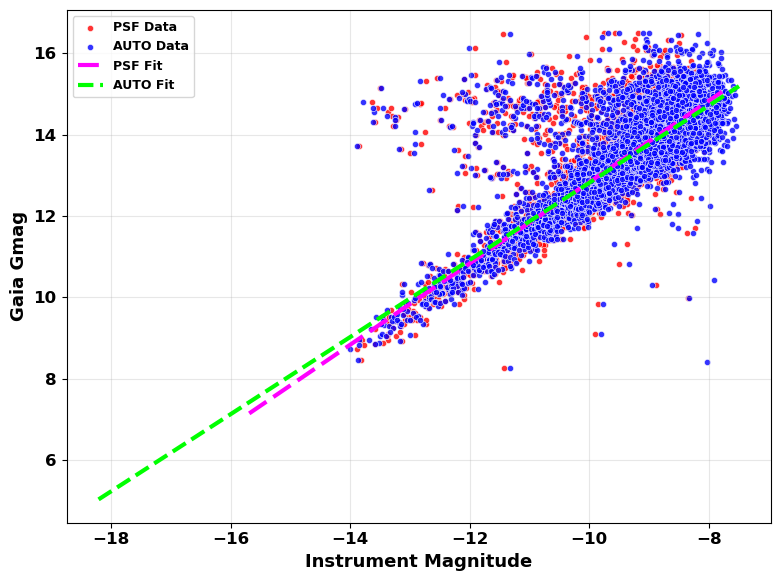}
    \includegraphics[height=0.45\linewidth, width=0.45\linewidth]{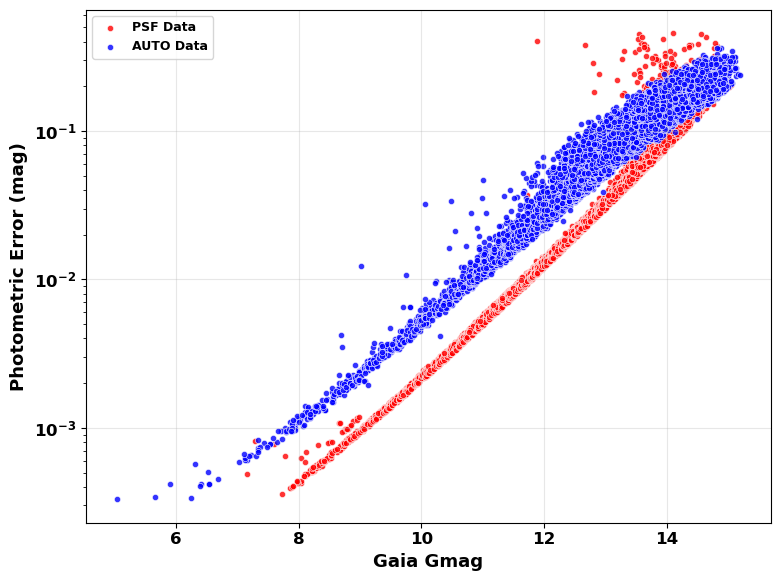}

    \caption{ Left panel: Fitting of instrument magnitudes to Gaia DR3 G-band magnitudes, with instrument magnitudes calculated by SExtractor using the formula -2.5$ log_{10}$ (Flux).  Right panel: Photometric magnitude errors as a function of Gaia G-band magnitudes.
}
    \label{fig:magvserror}
\end{figure*}

We extract the light curves for one day of data (168 observational images in total), to test the quality of the photometry. We cross-match the stars in the photometry catalogs with the Gaia DR3 catalog, based on their RA and Dec coordinates, for which matching radius is set as the average FWHM plus one $\sigma$ in each catalog. There are 52,580 stars observed more than 10 times. For the instrumental magnitudes in each catalog, we convert them to the G-band magnitudes used by Gaia DR3, using a simple linear transformation relationship (see the left panel of Figure~\ref{fig:magvserror}). The discrepancy between AUTO and PSF photometry methods arises from their fundamental principles: MAGERR\_AUTO estimates uncertainty through error propagation within an adaptive elliptical aperture. By contrast, MAGERR\_PSF derives its uncertainty directly from the covariance matrix of the PSF model fitting, reflecting the quality of the fit itself (see the right panel of Figure~\ref{fig:magvserror}). For 30-second exposures in the G-band, the 1.5~$\sigma$  detection limit reaches 15.00~mag, with photometric errors below 0.1~mag for the majority of stars brighter than 14.00~mag. This improves to better than 0.01~mag for most stars brighter than 11.00~mag and 12.00~mag when using the AUTO and PSF photometry methods, respectively (see the right panel of Figure~\ref{fig:magvserror}).

\section{Data Products}
\label{sect:products}

The prototype data products consist exclusively of processed and calibrated data files, with a photometry catalog for each science frame, following a standardized directory structure under the root directory \texttt{AT-Proto}. The data processing pipeline has systematically applied all necessary calibrations to the raw observational data, reduced to science-ready products. A date-stamped subdirectory (in the format \texttt{YYYY-MM-DD}) for each observation night contains these processed data products:

\begin{itemize}
        \item \texttt{master\_bias.fits} -- Combined bias frame (zero-second exposure)
        \item \texttt{master\_dark\_xs.fits} -- Combined dark frame with x seconds exposure
        \item \texttt{YYYY-MM-DD\_super\_science\_flat.fits} -- Master flat field image
        \item \texttt{R*T.fits} -- Fully processed science frames
        \item \texttt{R*T.cat} -- Photometry catalog
\end{itemize}

This dataset provides immediately usable science-ready products, with all calibration steps documented and in the header metadata. The FITS headers document the complete processing history through standardized keywords:
\begin{itemize}
    \item Processing flags (all set to \texttt{True}):
    \begin{itemize}
        \item \texttt{PIXFIXED} -- bad pixel correction
        \item \texttt{BLEEDING\_CORR} -- bleeding correction
        \item \texttt{SUBBIAS} -- master bias subtraction
        \item \texttt{SUBDARK} -- master dark subtraction
        \item \texttt{FLATCOR} -- flat fielding correction
        \item \texttt{BKGRDSUB} -- background sky subtraction
        \item \texttt{HISTORY} -- records of astrometric solutions (e.g., SCAMP processing)
    \end{itemize}
\end{itemize}

The catalogs of photometry consist of the following contents in each column:

\begin{itemize}
\item \texttt{XWIN\_IMAGE} -- Windowed position estimate along x
\item \texttt{YWIN\_IMAGE} -- Windowed position estimate along y
\item \texttt{ALPHAPSF\_J2000} -- Right ascension of the fitted PSF (J2000)
\item \texttt{DELTAPSF\_J2000} -- Declination of the fitted PSF (J2000)
\item \texttt{FLUX\_AUTO} -- Flux within a Kron-like elliptical aperture 
\item \texttt{FLUXERR\_AUTO} -- RMS error for AUTO flux
\item \texttt{MAG\_AUTO} -- Kron-like elliptical aperture magnitude
\item \texttt{MAGERR\_AUTO} -- RMS error for AUTO magnitude
\item \texttt{FLUX\_PSF} -- Flux from PSF-fitting
\item \texttt{FLUXERR\_PSF} -- RMS flux error for PSF-fitting
\item \texttt{MAG\_PSF} -- Magnitude from PSF-fitting 
\item \texttt{MAGERR\_PSF} -- RMS magnitude error from PSF-fitting
\item \texttt{FWHM\_WORLD} -- FWHM assuming a gaussian core (deg)
\item \texttt{ELLIPTICITY} -- 1~--~B\_IMAGE/A\_IMAGE
\item \texttt{SNR\_WIN} -- Gaussian-weighted SNR
\item \texttt{CLASS\_STAR} -- Star/Galaxy classifier output
\item \texttt{FLAGS} -- Extraction flags
\end{itemize}


\section{Summary and Prospective}
\label{sect:conclusion}
The ATSOP has successfully carried out a high-cadence and long-period optical observation experiment in Antarctica, using the 18-cm aperture AT-Proto telescope.  Deployed at Zhongshan Station, the system delivered robust performance under extreme environmental conditions. Throughout 2023, it accumulated 145 days of observational data, including a continuous observation campaign during the polar night (May 21 to July 15), and generated over 161,805 raw FITS images.

The 2023 dataset of the AT-Proto telescope is being released concurrently with the publication of this paper, and specifically includes two core product types: Reduced fits images and photometric catalogs. To convert raw observational data into scientifically usable products, the research team have developed a robust data reduction pipeline encompassing the four key stages of data preprocessing, instrumental effect correction, astrometric solution, and full-field stellar photometry. To date, the development of the Level-1 data processing pipeline for the prototype data has been finalized. This specialized pipeline addresses a range of instrumental artifacts, including bias, dark current, flat-field distortion, bad pixels, and shutter effects, and completes both positional calibration and flux calibration, ultimately yielding Level-1 processed image products. Key performance metrics of these products are astrometric precision of 1.9 and 1.7~arcsec in RA and Dec, a detection limit of 15.00~mag for 30-second exposures with a detection threshold of 1.5~$\sigma$, and photometric errors of less than 0.1~mag for the majority of stars brighter than 14.00~mag. These photometric errors improve to become better than 0.01~mag for most stars brighter than 11.0~mag and 12.0~mag when employing the AUTO and PSF photometry methods, respectively.

In future work, we plan to optimize the pipeline further. Subsequent research will focus on developing comprehensive flux calibration and de-trending of light curves from the processed data, facilitating in-depth investigations of variable stars, transient astrophysical phenomena, and related celestial events. These refined data products, together with the full pipeline specifications, will be published in an upcoming paper dedicated to our secondary data releases.

The ATSOP has successfully overcome the logistical and technical challenges of the harsh Antarctic environment, as demonstrated by the innovative AT-Proto telescope. The integrated design, featuring a temperature-controlled dome and trackless drift-scanning technology, has accomplished 248 consecutive days of non-stop operation, including during the polar night period. The success of AT-Proto provides a strong foundation for future projects, including the development of larger-aperture instruments and more sophisticated data processing methods. Building on this success, we plan to expand the observational sky coverage of ATSOP by deploying additional wide-field telescopes, with enhanced sensitivity for time-domain astronomy, while developing next-generation instruments and sophisticated data processing methods. These advancements will support key scientific goals, such as monitoring transient events, investigating high-inclination small bodies in the Solar System, studying stellar variability, and exploring dynamic cosmic phenomena. By continuing to refine both hardware and software, ATSOP is poised to become a cornerstone of southern hemisphere time-domain astronomy, contributing to global efforts in understanding the dynamic universe. 

\section*{acknowledgements}
This work is supported by the National Natural Science Foundation of China (NSFC) through the grants 12090040 and 12090042, science research grants from the China Manned Space Project (NO. CMS-CSST-2025-A19), the Shanghai Academic/Technology Research Leader Program, and sponsorship from the Natural Science Foundation of Xinjiang Uygur Autonomous Region (2023D01A12). The research is partly supported by the Operation, Maintenance and Upgrading Fund for Astronomical Telescopes and Facility Instruments, budgeted from the Ministry of Finance of China (MOF) and administrated by the Chinese Academy of Sciences (CAS).

This research has made use of the VizieR catalogue access tool, CDS, Strasbourg, France, and also ccdproc, an Astropy package for image reduction (\citeauthor{2017zndo...1069648C}). This work made use of Astropy\footnote{http://www.astropy.org}: a community-developed core Python package and an ecosystem of tools and resources for astronomy\citep{astropy:2022}.

\section*{Author Contributions}
Hubiao Niu developed the data processing pipeline, finalized the data products for AT-Proto in 2023 and wrote the paper. Jing Zhong assisted with data processing and wrote the paper. Yu Zhang, Jianchun Shi, Rui Rong, Jinzhong Liu, Shiyin Shen, Zhenghong Tang, and Dan Zhou
 participated in the organization of data and discussions on the data processing pipeline, and provided suggestions for the data processing work. All authors read and approved the final manuscript.

\bibliography{ati}{}
\bibliographystyle{aasjournal}



\end{document}